\documentclass[sigplan,screen]{acmart}

\usepackage{hyperref} % for clickable citations   
\usepackage{subcaption}

\AtBeginDocument{%
  }

\setcopyright{acmlicensed}
\copyrightyear{2018}
\acmYear{2018}
\acmDOI{XXXXXXX.XXXXXXX}
\acmConference[Conference acronym 'XX]{Make sure to enter the correct
  conference title from your rights confirmation email}{June 03--05,
  2018}{Woodstock, NY}
\acmISBN{978-1-4503-XXXX-X/2018/06}

\copyrightyear{2025}
\acmYear{2025}
\setcopyright{cc}
\setcctype{by-nc}
\acmConference[HT 2025]{Proceedings of the 36th ACM Conference on Hypertext and Social Media}{September 15--18, 2025}{Chicago, IL, USA}
\acmBooktitle{Proceedings of the 36th ACM Conference on Hypertext and Social Media (HT 2025), September 15--18, 2025, Chicago, IL, USA}
\acmDOI{10.1145/3720553.3746686}
\acmISBN{979-8-4007-1534-1/2025/09} 

\begin{document}

\title[Comprehensive Privacy Risk Assessment in Social Networks]{Comprehensive Privacy Risk Assessment in Social Networks Using User Attributes Social Graphs and Text Analysis}

% \author{Md Jahangir Alam}
% % \authornote{Both authors contributed equally to this research.}
% \email{malam10@miners.utep.edu}
% \orcid{0009-0005-8731-7354}
% % \author{G.K.M. Tobin}
% \affiliation{%

\author{Md Jahangir Alam}
\affiliation{%
  \institution{The University of Texas at El Paso}
  \city{El Paso}
  \state{Texas}
  \country{USA}}
\email{malam10@miners.utep.edu}

\author{Ismail Hossain}
\affiliation{%
  \institution{The University of Texas at El Paso}
  \city{El Paso}
  \state{Texas}
  \country{USA}}
\email{ihossain@miners.utep.edu}

\author{Sai Puppala}
\affiliation{%
 \institution{Southern Illinois University Carbondale}
 \city{Carbondale}
 \state{Illinois}
 \country{USA}}
\email{sai.puppala@siu.edu}

\author{Sajedul Talukder}
\affiliation{%
  \institution{The University of Texas at El Paso}
  \city{El Paso}
  \state{Texas}
  \country{USA}}
\email{stalukder@utep.edu}

\renewcommand{\shortauthors}{Md Jahangir et al.}

\begin{abstract}
The rise of social networking platforms has amplified privacy threats as users increasingly share sensitive information across profiles, content, and social connections. We present a \textit{Comprehensive Privacy Risk Scoring (CPRS)} framework that quantifies privacy risk by integrating user attributes, social graph structures, and user-generated content. Our framework computes risk scores across these dimensions using sensitivity, visibility, structural similarity, and entity-level analysis, then aggregates them into a unified risk score. We validate CPRS on two real-world datasets: the SNAP Facebook Ego Network (4,039 users) and the Koo microblogging dataset (1M posts, 1M comments). The average CPRS is 0.478 with equal weighting, rising to 0.501 in graph-sensitive scenarios. Component-wise, graph-based risks (mean 0.52) surpass content (0.48) and profile attributes (0.45). High-risk attributes include Email, Date of Birth, and Mobile Number. Our user study with 100 participants shows 85\% rated the dashboard as clear and actionable, confirming CPRS’s practical utility. This work enables personalized privacy risk insights and contributes a holistic, scalable methodology for privacy management. Future directions include incorporating temporal dynamics and multimodal content for broader applicability.
\end{abstract}







\maketitle

\section{Introduction}
With the rapid growth of social networks, users are increasingly exposed to privacy risks. Sensitive data shared on these platforms including posts, comments, and profile attributes which creates significant vulnerabilities that attackers can exploit. Prior studies emphasize the need to quantify such risks. Liu and Terzi~\cite{LiuTerzi2009} introduced visibility-based privacy risk quantification in social graphs, while Li et al.~\cite{li2024method} demonstrated the role of sensitive entity detection in user-generated content for enhancing privacy assessments. However, existing methods often focus on isolated factors such as user attributes, content, or graph structures~\cite{li2024method, JehWidom2002, Page1998, zhao2019ranking}, missing a comprehensive view that integrates these dimensions. Recent findings show that nearly 75\% of users share personal information publicly or with limited settings, leaving them vulnerable to various attacks. Graph-based threats like sybil attacks, link prediction, and de-anonymization~\cite{Backstrom2007} exploit user relationships to infer private information.

Profile attributes such as names, emails, phone numbers, and locations are frequently used in social engineering attacks, phishing, and identity theft~\cite{LiuTerzi2009}. Sensitive entities in posts and comments further expose users to context inference and de-anonymization risks~\cite{li2024method, hua2020xref}. Attackers can also exploit graph structures to predict hidden relationships~\cite{Backstrom2007}, or infer sensitive locations from geotagged content and behavioral patterns~\cite{ghinita2016protecting, zhao2019ranking}.

Effective privacy risk assessment is essential for enabling users and platform operators to make informed privacy decisions. Users often lack insight into how their profile settings, social connections, and shared content together influence their privacy exposure. A unified framework to quantify these risks can support the creation of privacy dashboards, automated recommendations, and transparency tools that enhance user trust and regulatory compliance. Despite rising concerns, few existing approaches capture both structural and content-based privacy risks. To address this, we propose a \textit{Comprehensive Privacy Risk Scoring (CPRS)} framework that integrates user attribute sensitivity, social graph structure, and content-based exposure into a unified model. We evaluate our approach using the SNAP~\cite{jure2014snap} and Koo~\cite{mekacher2024koo} datasets, representing social graph and microblogging platforms.

The goal of our privacy risk score is twofold: (1) to raise user awareness about potential privacy risks arising from oversharing of sensitive information, and (2) to provide actionable guidance for adjusting privacy settings and content exposure. By quantifying risk across multiple dimensions, our framework enables more informed privacy choices.

\noindent \textbf{Research Questions.}
In this paper we aim to address the following research questions:

\textbf{RQ1:} How to quantify privacy risk by integrating profile attributes, social graph structures, and user-generated content?

\textbf{RQ2:} What are the relative contributions of these dimensions to the overall privacy exposure of social network users?

\textbf{RQ3:} Can the proposed framework generate actionable insights to help users manage and reduce their privacy risks?

\textbf{RQ4:} How effective is the framework when applied to real-world datasets and evaluated through user studies?

\noindent \textbf{Key Contributions.}
Our main contributions are as follows: We propose \textit{CPRS}, a holistic framework that integrates attribute-, graph-, and content-based privacy risk factors. It computes risk using attribute sensitivity and visibility, structural analysis via SimRank and PageRank, and entity extraction from user-generated content. We evaluate CPRS on two large-scale datasets, SNAP and Koo, demonstrating its applicability across diverse network structures and content types. Finally, we validate its usability through a user study with 100 participants, showing strong acceptance and effectiveness in delivering actionable privacy recommendations.

\section{Related Work}
\subsection{Sensitive Information Related Privacy Attacks}
Social network users face a range of privacy threats from different attack vectors. One common method exploits publicly visible profile attributes—like name, email, phone number, and location—targeted in social engineering, phishing, and identity theft attacks. Liu and Terzi~\cite{LiuTerzi2009} showed that weak privacy settings heighten such risks. User-generated content, such as posts and comments containing names, locations, or financial data, can lead to de-anonymization and context inference. Li et al.~\cite{li2024method} and Hua et al.~\cite{hua2020xref} showed how sensitive entities can be extracted and linked to external data to infer private details. Social graphs also face structural attacks. Backstrom et al.~\cite{Backstrom2007} showed that graph structures can reveal hidden links or re-identify anonymized users. Likewise, geotagged posts and check-ins can expose sensitive locations such as home or workplace. Ghinita et al.~\cite{ghinita2016protecting} demonstrated that behavior patterns in location-based networks can be analyzed to infer such locations. These attack models underscore the need for comprehensive risk assessment frameworks.

\subsection{Attribute Visibility-Based Privacy Risk}
Liu and Terzi~\cite{LiuTerzi2009} proposed visibility weighting methods that assign binary weights to user privacy settings, with higher weights for publicly visible attributes. While their approach is simple and scalable, it does not consider the sensitivity of user-generated content, limiting its ability to capture nuanced privacy risks. Blanco-Aza et al.~\cite{BlancoAza_Daniel_Comprehensive} introduced a comprehensive AI-driven privacy risk assessment framework that incorporates audience size and visibility factors to quantify privacy risks more precisely. Similarly, De and Imine~\cite{de2017privacy} proposed a user-centric privacy scoring mechanism that evaluates attribute visibility in social graphs to assess disclosure risks based on exposure to different audiences.

\subsection{Content-Based Risk via Entity Extraction}
Entity extraction is central to social media privacy research. Li et al.~\cite{li2024method} proposed methods for extracting sensitive entities from posts and comments, though limited by predefined entity types. Recent multimodal approaches enhance entity and relation recognition by combining text and visual cues. Moon et al.~\cite{moon2018mner} introduced a modality-attentive BiLSTM-CRF model using visual features to improve NER. Yuan et al.~\cite{yuan2024jmere} proposed a cross-modal prompting framework for few-shot joint extraction from text-image pairs. Wu et al.~\cite{wu2023mcgmner} presented MCG-MNER, a generative model leveraging hierarchical visual features and instructions. Zhang et al.~\cite{zhang2023multigranularity} developed a multi-granularity model with hierarchical attention and gating. Li et al.~\cite{li2024method} extended NER to identify identity groups and organizations in comments. Hua et al.~\cite{hua2020xref} focused on entity linking in Chinese news comments, highlighting persons, locations, and events. These works underscore the role of entity-level sensitivity in evaluating content-based privacy risks.

\subsection{Graph-Based Privacy Risk}
Graph-based methods have been used to evaluate privacy risks stemming from user connections and structural positioning. Zhao et al.~\cite{zhao2019ranking} introduced a motif-based PageRank framework to assess user importance and structural privacy risks, capturing higher-order relations in social graphs. However, their method faces scalability challenges on large networks. SimRank~\cite{JehWidom2002} and PageRank~\cite{Page1998} have also been employed to measure structural similarity and user influence, though these methods typically overlook content-based exposures. Backstrom et al.~\cite{Backstrom2007} highlighted structural vulnerabilities such as link prediction and de-anonymization attacks, demonstrating the importance of structural privacy assessments. While attribute-based~\cite{LiuTerzi2009}, content-based~\cite{li2024method, yuan2024jmere, zhang2023multigranularity}, and graph-based~\cite{zhao2019ranking} methods have shown promise individually, few works combine these dimensions into a unified framework. Existing studies often fail to account for the interplay between visibility, content sensitivity, and network structure, limiting their applicability in real-world scenarios. This gap motivates the development of integrated models, such as the one we propose in this paper, that comprehensively assess privacy risks across multiple user data dimensions.

\section{Methodology}\label{methodology}
In this section, we outline our methodology to address \textbf{RQ1–RQ3}, presenting how we compute attribute-based, graph-based, and content-based privacy risk scores and integrate them into a unified, weighted Comprehensive Privacy Risk Score (CPRS) using a weighted sum of the three components:

\begin{equation}
\label{eq:weighted_sum_of_three_components}
    CPRS = w_1 \cdot APRS + w_2 \cdot SGPRS + w_3 \cdot CBPRS
\end{equation}

Here, \( w_1, w_2, \) and \( w_3 \) represent the relative importance of each component. We explore three weighting strategies (Table~\ref{tab:privacy_risk_weights}): equal weighting as a baseline, content-focused for microblog platforms, and graph-focused for social graphs like Facebook. To ensure systematic justification of the weights, we apply the Analytic Hierarchy Process (AHP)~\cite{saaty1987ahp}.

\begin{table}[ht]
\centering
\caption{Weights for Privacy Risk Components}
\label{tab:privacy_risk_weights}
\resizebox{0.99\linewidth}{!}{
\begin{tabular}{lccc}
\toprule
\textbf{Scenario} & \textbf{APRS ($w_1$)} & \textbf{SGPRS ($w_2$)} & \textbf{CBPRS ($w_3$)} \\
\midrule
Equal Importance              & 0.33 & 0.33 & 0.33 \\
Content-Focused Platform      & 0.20 & 0.30 & 0.50 \\
Social Graph-Focused Platform & 0.10 & 0.60 & 0.30 \\
\bottomrule
\end{tabular}
}
\end{table}

\subsection{Framework Overview}
\subsubsection{Attribute-Based Privacy Risk Score (APRS)}
APRS measures the sensitivity and visibility of user attributes like age, gender, location, and political views etc. All the user profile attributes we consider in our analysis are: \textit{Mobile, Email, Gender, Pronoun, Date of Birth, Relationship Status, From Location, Lives In Location, School}, and \textit{Workplace}. We infer visibility score of an attribute from the privacy settings of that attribute. We calculate visibility scores based on audience size of the attribute based on privacy settings of that attribute.

\subsubsection{Social Graph Privacy Risk Score (SGPRS)}
SGPRS quantifies a user’s privacy risk in the social graph by evaluating structural similarity and node importance. Structural similarity is measured using the SimRank algorithm, which identifies users with similar graph positions—higher similarity indicates increased risk of profile inference. Node importance is assessed using the PageRank algorithm, where higher importance implies greater visibility and vulnerability. These two factors are combined as a weighted sum, with $w_{sim}$ and $w_{imp}$ denoting the respective weights for similarity and importance. The resulting SGPRS captures the user's relational risk and informs graph-based privacy mitigation strategies.

\subsubsection{Content-Based Privacy Risk Score (CBPRS)}
To compute CBPRS, we analyze user posts and comments using entity extraction techniques to identify sensitive entities. The risk score is based on both the quantity and sensitivity of these entities—more numerous or sensitive entities increase the score. Post and comment risks are then aggregated for a final score. Even when users knowingly share entities like names or locations, they remain vulnerable due to potential cross-referencing. Combining such details with external data can reveal private information. Prior work shows even limited public data can enable de-anonymization or inference~\cite{li2019matching}, emphasizing the importance of entity-level sensitivity in privacy assessment. In the following sections we describe how we compute the components scores of the CPRS.

\section{Attribute Based Privacy Risk Score}
Let \( A = \{a_1, a_2, \dots, a_n\} \) represent the set of \( n \) profile attributes, and \( U = \{u_1, u_2, \dots, u_m\} \) represent the \( m \) users. Each user \( u_j \) (\( j = 1, 2, \dots, m \)) has a profile defined by a set of attributes \( \{a_{1j}, a_{2j}, \dots, a_{nj}\} \). For each attribute \( a_i \) (\( i = 1, 2, \dots, n \)) let \( S(a_i) \) be the sensitivity score of attribute \( a_i \), indicating how sensitive the attribute is to privacy risks and \( V(a_{i}) \) be the visibility score of attribute \( a_i \) for user \( u_j \), representing how publicly the attribute is exposed based on privacy settings.

\subsection{Attribute Sensitivity Calculation}
Let \( f(a_i) \) denote the frequency of attribute \( a_i \) across all users. We utilize the sensitivity \( S(a_i) \) calculation formula which is a function of attributes frequency, which is the Attribute Frequency Inverse User Frequency (AFIUF) \cite{gong2022privacy}:

\begin{equation}
\label{eq:afiuf}
    S(a_i) = \frac{f(a_i)}{\log_2\left(\frac{m}{f(a_i)} + 1\right)}
\end{equation}

Rare attributes (e.g., unique political views or uncommon workplaces) are assigned higher sensitivity scores due to their increased potential to uniquely identify users. Such attributes have higher entropy and stand out within datasets, making them attractive for profiling or inference attacks. The AFIUF formula captures this by assigning greater sensitivity to infrequent attributes, reflecting their elevated privacy risk.

\subsection{Attribute Visibility Calculation}
The visibility of an attribute \( a_{ij} \) for user \( u_j \) depends on its assigned privacy setting. We define visibility \( V(a_{ij}) \) as:

\begin{equation}
\label{eq:attribute_visibility}
    V(a_{ij}) = \frac{\text{Audience Size for } \text{P}(a_{ij})}{A_{\max}}
\end{equation}

Here, \( \text{P}(a_{ij}) \) is the privacy setting (e.g., ``Public", ``Friends-Only", ``Only Me"), and \( A_{\max} \) is the maximum audience size—equal to the network size when the setting is ``Public" For example, in a 1 million-user network, $A_{\max} = 1{,}000{,}000$, giving a visibility of 1.0 for public attributes. ``Friends-Only" visibility is computed as friend count over $A_{\max}$, while ``Only Me" is assigned a fixed visibility of 0.1 to account for minimal yet non-zero risk (e.g., account compromise). We assign privacy setting weights based on visibility and potential for unauthorized access. ``Public" attributes carry highest risk, ``Friends-Only" settings reflect moderate exposure within a possibly untrustworthy circle, and ``Only Me" settings carry minimal risk, though not entirely immune to breaches. These visibility values are essential for quantifying attribute-level privacy risk.

\subsection{Privacy Risk Score Calculation}
The privacy risk \( R(a_{ij}) \) for an attribute \( a_{ij} \) of user \( u_j \) is computed as the product of sensitivity and visibility:

\begin{equation}
\label{eq:attribute_based_privacy_risk_score}
    R(a_{ij}) = S(a_i) \cdot V(a_{ij})
\end{equation}

The total profile-based privacy risk score \( R(u_j) \) for user \( u_j \) is then the sum of risks across all \( n \) attributes:

\begin{equation}
\label{eq:profile_based_attribute_based_privacy_risk_score}
    R(u_j) = \sum_{i=1}^{n} R(a_{ij}) = \sum_{i=1}^{n} S(a_i) \cdot V(a_{ij})
\end{equation}

This score provides a comprehensive measure of the privacy risk posed by the user’s profile attributes, accounting for both the inherent sensitivity of the attributes and their visibility to others.

\section{Social Graph Based Privacy Risk Score}
The Social Graph Privacy Risk Score (SGPRS) quantifies the privacy risk associated with a user's position and interaction patterns in a social network graph. The score is based on two primary factors: \textit{structural similarity} and \textit{user importance}.

Let \( G = (V, E) \) be the social graph, where \( V = \{u_1, u_2, \dots, u_n\} \) represents the set of \( n \) users (nodes), and \( E \) represents the connections (edges) between users.
\( A \) be the adjacency matrix of \( G \), where \( A_{ij} = 1 \) if there is a connection between \( u_i \) and \( u_j \), and \( A_{ij} = 0 \) otherwise. \( S(u_i, u_j) \) be the structural similarity score between users \( u_i \) and \( u_j \), computed using the SimRank algorithm. \( P(u_i) \) be the PageRank score of user \( u_i \), indicating their importance in the graph.

\subsection{Structural Similarity}
The structural similarity quantifies how closely connected a user is to others with similar sharing behaviors. Users are at higher risk if their neighbors also share sensitive information.

\subsubsection{SimRank Score}
The SimRank score \( S(u_i, u_j) \) measures the similarity between two users \( u_i \) and \( u_j \) using the following equation:

\begin{equation}
\label{eq:simrank_score}
    S(u_i, u_j) = \frac{C}{|N(u_i)| \cdot |N(u_j)|} \sum_{v_k \in N(u_i)} \sum_{v_l \in N(u_j)} S(v_k, v_l)
\end{equation}
where \( N(u_i) \) is the set of neighbors of \( u_i \). \( C \) is the decay factor (\( 0 < C < 1 \)). \( S(v_k, v_l) \) is the similarity between neighbors \( v_k \) and \( v_l \).

\subsubsection{Normalized Structural Similarity Risk}
The structural risk score for a user \( u_i \) is:

\begin{equation}
\label{eq:normalized_structural_similarity_score}
    R_{\text{struct}}(u_i) = \frac{1}{|N(u_i)|} \sum_{u_j \in N(u_i)} S(u_i, u_j) \cdot R_{\text{neighbor}}(u_j)
\end{equation}

where \( R_{\text{neighbor}}(u_j) \) is the privacy risk of neighbor \( u_j \). We use the ABPRS value for \( R_{\text{neighbor}}(u_j) \) in our experiment.

This score reflects user risk by aggregating neighbors’ risks weighted by structural similarity and normalizing by neighbor count. High scores indicate proximity to high-risk or similar neighbors, while low scores suggest safer network positions.

\subsection{User Importance}
User importance quantifies the influence and visibility of a user in the graph. Highly central users are more vulnerable to privacy breaches due to their reach.

\subsubsection{PageRank Score}
The PageRank algorithm ~\cite{brin1998anatomy}, developed by Larry Page and Sergey Brin in 1996, is foundational to Google's search technology. The formula for calculating the PageRank score \( P(u_i) \) of a user \( u_i \) is:

\begin{equation}
\label{eq:pagerank_score}
    P(u_i) = (1 - d) + d \sum_{u_j \in N(u_i)} \frac{P(u_j)}{|N(u_j)|}
\end{equation}

where \( d \) represents the damping factor (where \( 0 < d < 1 \)), and \( |N(u_j)| \) denotes the number of neighbors of user \( u_j \).

\subsubsection{Normalized Importance Risk}
This score quantifies a user's privacy risk based on their prominence within a social network. This metric is derived by normalizing the user's PageRank score \( P(u_i) \) against the highest PageRank score in the network:

\begin{equation}
\label{eq:normalized_pagerank_score}
    R_{\text{imp}}(u_i) = \frac{P(u_i)}{\max_{u_k \in V} P(u_k)}
\end{equation}

where \( P(u_i) \) represents the PageRank score of user \( u_i \), and \( \max_{u_k \in V} P(u_k) \) denotes the maximum PageRank score among all users in the network. By normalizing in this manner, the importance risk score \( R_{\text{imp}}(u_i) \) ranges between 0 and 1, facilitating straightforward comparisons across users.

\subsection{Aggregated Risk}
The Social Graph Privacy Risk Score (SGPRS) for user \( u_i \) combines structural similarity and user importance:

\begin{equation}
\label{eq:social_graph_based_aggregated_risk}
    SGPRS(u_i) = w_1 \cdot R_{\text{struct}}(u_i) + w_2 \cdot R_{\text{imp}}(u_i)
\end{equation}

Weights $w_1$ and $w_2$ are set based on correlation with attribute-based risk. Structural similarity shows stronger correlation (0.68) than importance (0.55), so we assign it higher weight. A high $R_{\text{struct}}(u_i)$ means the user resembles high-risk neighbors, increasing co-disclosure risk. A high $R_{\text{imp}}(u_i)$ indicates centrality, implying wider potential exposure. SGPRS combines these factors into a comprehensive social graph risk score.

\section{Content-Based Privacy Risk Score}
\subsection{Post-Level Privacy Risk}
Let \( U = \{u_1, u_2, \dots, u_m\} \) be the set of users and lets denote the set of posts by \( P = \{p_1, p_2, \dots, p_N\} \) made by user \( u_j \), where each post \( p_i \) includes a comment set \( C(p_i) \), a sensitivity score \( S(p_i) \), and a visibility score \( V(p_i) \) based on its audience. We compute the privacy risk for \( p_i \) by extracting sensitive entities from text using SpaCy, which detects 18 entity types (e.g., \textit{PERSON, LOCATION, DATE, ORGANIZATION, GPE, EMAIL, MONEY}). These entities inform content-based risk scores via aggregated sensitivity and visibility.

Let the post \( p_i \) contain text component \( T(p_i) \) with extracted entities \( E_T(p_i) = \{e_1, e_2, \dots, e_{n_T}\} \). Each entity \( e_k \) is assigned a sensitivity score \( S(e_k) \), based on its type. The total sensitivity score of the post is:

\begin{equation}
\label{eq:sensitivity_score_of_user_post}
    S(p_i) = \sum_{e_k \in E_T(p_i)} S(e_k)
\end{equation}

The visibility score \( V(p_i) \) depends on the post’s privacy setting (e.g., ``Public", ``Friends-Only", ``Only Me"). The final privacy risk score for the post is:

\begin{equation}
\label{eq:visibility_score_of_post}
    R(p_i) = S(p_i) \cdot V(p_i)
\end{equation}

Here, \( R(p_i) \) reflects the overall privacy risk, combining entity sensitivity and visibility.

\subsection{Comment-Level Privacy Risk}
User comments on a post can reveal additional sensitive information and elevate the post’s overall privacy risk. Let \( C(p_i) = \{c_1, c_2, \dots, c_M\} \) be the set of textual comments on post \( p_i \). For each comment \( c_k \), we extract sensitive entities \( E(c_k) = \{e_1, e_2, \dots, e_n\} \), each assigned a sensitivity score \( S(e_k) \) based on its type. The sensitivity score of a comment is computed as:

\begin{equation}
\label{eq:comment_sensitivity_score}
    S(c_k) = \sum_{e_k \in E(c_k)} S(e_k)
\end{equation}

The visibility score of \( c_k \) is inherited from the associated post, i.e., \( V(p_i) \). The privacy risk for each comment is:

\begin{equation}
\label{eq:comment_visibility_score}
    R(c_k) = S(c_k) \cdot V(p_i)
\end{equation}

The total privacy risk from all comments is then aggregated as:

\begin{equation}
\label{eq:total_comment_privacy_risk}
    R(C(p_i)) = \sum_{k=1}^{M} R(c_k)
\end{equation}

This cumulative score \( R(C(p_i)) \) reflects the overall contribution of comments to the privacy risk of the post.

\subsection{Overall Content Based Privacy Risk}
The total privacy risk of a post \( p_i \), including its content and comments, is calculated as:

\begin{equation}
\label{eq:combined_post_and_comment_privacy_risk}
    R_{\text{Total}}(p_i) = R(p_i) + R(C(p_i))
\end{equation}

Here, \( R(p_i) \) is the risk from post content, and \( R(C(p_i)) \) is the cumulative risk from comments. This combined score \( R_{\text{Total}}(p_i) \) captures the overall exposure due to both post and comment content, accounting for scenarios where comments reveal additional sensitive information.

To compute the user-level content-based privacy risk for user \( u_j \), we aggregate the total risks across all their posts:

\begin{equation}
\label{eq:user_level_content_based_privacy_risk_score}
    CBPRS(u_j) = \sum_{i=1}^{N} R_{\text{Total}}(p_i)
\end{equation}

where \( N \) is the number of posts by \( u_j \), and \( R_{\text{Total}}(p_i) \) is the total privacy risk for each post.

\section{Comprehensive Privacy Risk Score}
To get the Comprehensive Privacy Risk Score (CPRS) we combine the results from Attribute Based Privacy Risk Score (APRS), Social Graph Based Privacy Risk Score (SGPRS), and Content Based Privacy Risk Score (CBPRS). The combined score is calculated using the Equation~\ref{eq:weighted_sum_of_three_components}.
\begin{figure*}[t]
    \centering
    \begin{subfigure}[t]{0.32\textwidth}
        \centering
        \includegraphics[width=\linewidth]{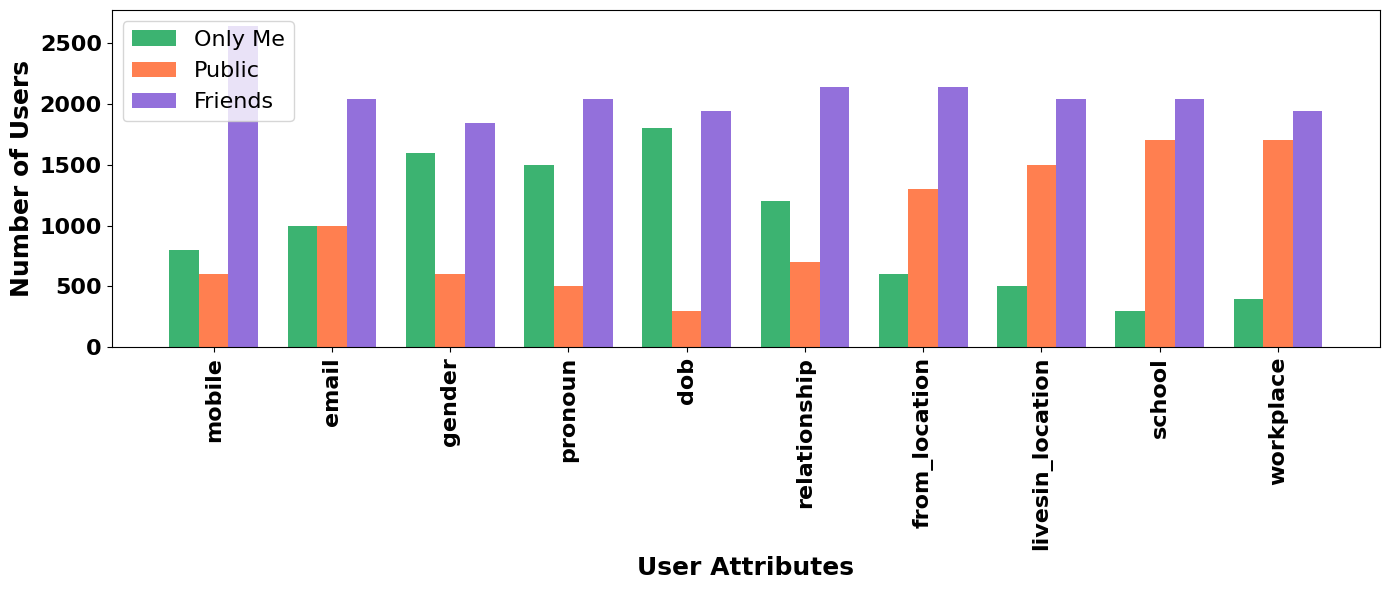}
        \caption{Privacy Settings Distribution}
        \Description{Bar chart showing distribution of privacy settings for each attribute.}
        \label{fig:privacy_settings_distribution}
    \end{subfigure}
    \hfill
    \begin{subfigure}[t]{0.32\textwidth}
        \centering
        \includegraphics[width=\linewidth]{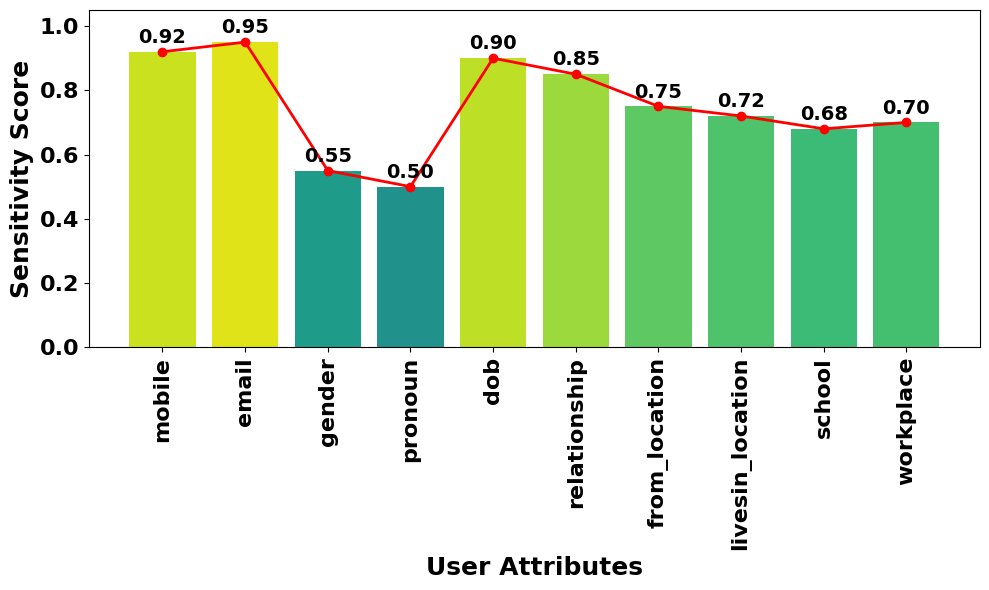}
        \caption{Attribute Sensitivity Scores}
        \Description{Bar chart showing sensitivity scores across attributes.}
        \label{fig:attribute_sensitivity_scores_across_dataset}
    \end{subfigure}
    \hfill
    \begin{subfigure}[t]{0.32\textwidth}
        \centering
        \includegraphics[width=\linewidth]{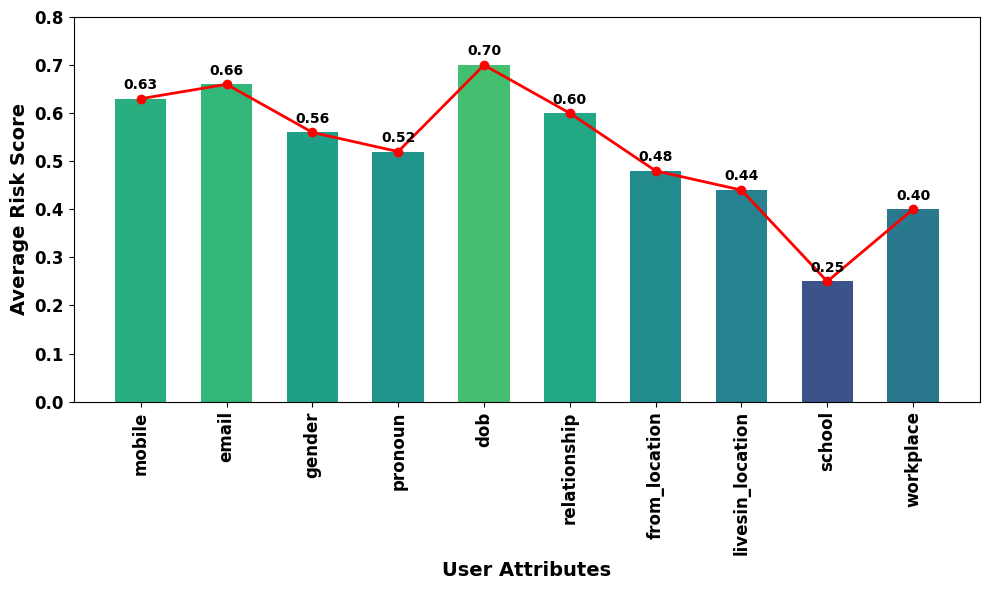}
        \caption{Average Risk Scores}
        \Description{Bar chart showing average privacy risk scores for user attributes.}
        \label{fig:average_risk_scores_per_attribute}
    \end{subfigure}
    \caption{Comparison of attribute-level privacy dimensions across the dataset: (a) Distribution of privacy settings (Public, Friends, Only Me) applied to each attribute; (b) Sensitivity scores reflecting inherent privacy importance of attributes; and (c) Average privacy risk scores computed by combining sensitivity and visibility.}
    \label{fig:attribute_risk_visuals}
\end{figure*}

\section{Experimental Setup}
We evaluate the Comprehensive Privacy Risk Score (CPRS) framework using the Stanford Facebook Ego Dataset, which includes 4,039 users and 88,234 connections. As this dataset lacks profile attributes, we generate synthetic ones using probabilistic distributions. For instance, age and gender follow distributions reflective of diverse online populations, and visibility settings mirror common configurations (e.g., ``Only Me" for Email, ``Friends Only" for School). To enhance realism, we incorporate homophily by assigning similar attributes to connected users, simulating real-world network behavior.

For content-based risk evaluation, we use the Koo dataset, containing 72M posts and 75M comments. We ensure temporal diversity by dividing the data into monthly intervals and uniformly sampling an equal number from each, resulting in 1M posts and 1M comments. These were randomly assigned across the 4,039 users to simulate realistic activity timelines for CPRS computation.
Figure~\ref{fig:privacy_settings_distribution} shows the distribution of privacy settings across the synthetic attributes.

\section{Experimental Results}
In this section, we evaluate our framework on real-world datasets, addressing \textbf{RQ2 and RQ4} by analyzing the contributions of different risk dimensions and demonstrating the overall effectiveness of the CPRS framework. Table~\ref{tab:privacy_risk_distribution} reports the distribution of normalized privacy risks across Attribute-Based (APRS), Social Graph-Based (SGPRS), and Content-Based (CBPRS) components. APRS ranges from 0.12 to 0.89 (mean 0.45), SGPRS from 0.08 to 0.91 (mean 0.52), and CBPRS from 0.10 to 0.87 (mean 0.48), reflecting variability in attribute sensitivity, network structure, and content exposure. 

\begin{table}[h]
\centering
\caption{Distribution of Privacy Risks}
\label{tab:privacy_risk_distribution}
\resizebox{0.9\linewidth}{!}{
\begin{tabular}{lccc}
\toprule
\textbf{Component} & \textbf{Min} & \textbf{Mean} & \textbf{Max} \\
\midrule
Attribute-Based Privacy Risk (APRS) & 0.12 & 0.45 & 0.89 \\
Social Graph Privacy Risk (SGPRS)   & 0.08 & 0.52 & 0.91 \\
Content-Based Privacy Risk (CBPRS)  & 0.10 & 0.48 & 0.87 \\
\bottomrule
\end{tabular}
}
\end{table}

Table~\ref{tab:cprs_results} presents the overall Comprehensive Privacy Risk Score (CPRS) computed using the weighting strategies defined in Equation~\ref{eq:weighted_sum_of_three_components}. Equal weighting (0.33 each) yields a CPRS of 0.478, content-focused weighting (0.2, 0.3, 0.5) increases it to 0.486, and graph-focused weighting (0.1, 0.6, 0.3) results in the highest CPRS of 0.501. These results demonstrate the flexibility of the CPRS framework in capturing platform-specific privacy risk variations.

\begin{table}[ht]
\centering
\caption{Comprehensive Privacy Risk Score (CPRS) under Different Weighting Methods}
\label{tab:cprs_results}
\resizebox{1.0\linewidth}{!}{
\begin{tabular}{lccccccc}
\toprule
\textbf{Weighting Method} & \textbf{\(w_1\)} & \textbf{\(w_2\)} & \textbf{\(w_3\)} & \textbf{APRS} & \textbf{SGPRS} & \textbf{CBPRS} & \textbf{CPRS} \\
\midrule
Equal Importance              & 0.33 & 0.33 & 0.33 & 0.45 & 0.52 & 0.48 & 0.478 \\
Content-Focused Platform      & 0.20 & 0.30 & 0.50 & 0.45 & 0.52 & 0.48 & 0.486 \\
Social Graph-Focused Platform & 0.10 & 0.60 & 0.30 & 0.45 & 0.52 & 0.48 & 0.501 \\
\bottomrule
\end{tabular}
}
\end{table}

\textbf{Attribute Sensitivity and Risk Scores.} Figures~\ref{fig:attribute_sensitivity_scores_across_dataset} and~\ref{fig:average_risk_scores_per_attribute} present the sensitivity and average risk scores for user attributes across the dataset. The top five sensitive attributes are Email, Mobile Number, Date of Birth, Relationship Status, and From Location, indicating strong contributions to overall privacy risk. Attributes like Date of Birth, Email, and Mobile also show high average risk scores due to their sensitivity and broad visibility, suggesting the need for stricter privacy settings. In contrast, attributes such as School and Workplace are less sensitive and exhibit lower risk.

\section{User Study}\label{user_study}
\begin{figure*}[t]
    \centering
    \begin{subfigure}[t]{0.32\textwidth}
        \centering
        \includegraphics[width=\linewidth]{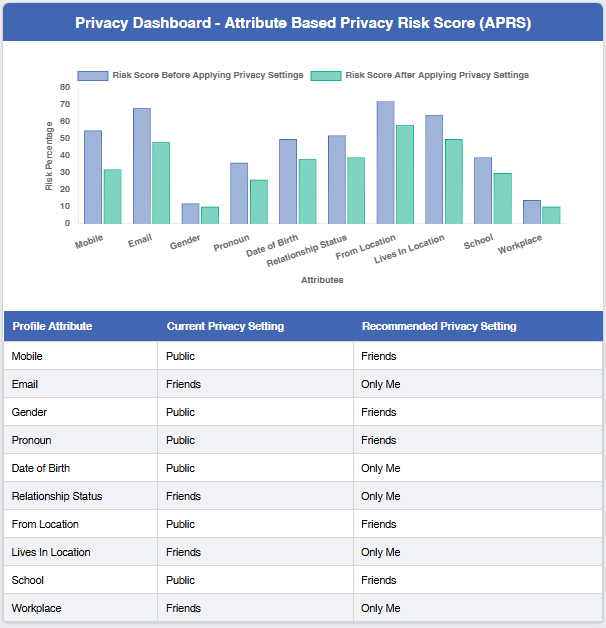}
        \caption{Attribute-Based Risk UI}
        \Description{Dashboard interface for displaying attribute-level privacy risk and recommendations.}
        \label{fig:dashboard_attribute_risk_score_ui}
    \end{subfigure}
    \hfill
    \begin{subfigure}[t]{0.32\textwidth}
        \centering
        \includegraphics[width=\linewidth]{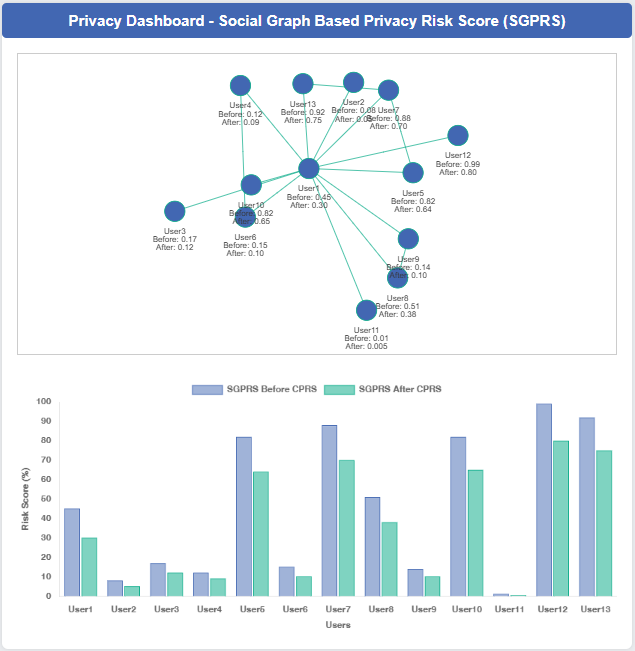}
        \caption{Graph-Based Risk UI}
        \Description{Dashboard showing social graph-based privacy risk using network analysis.}
        \label{fig:dashboard_social_graph_based_privacy_risk_score_ui}
    \end{subfigure}
    \hfill
    \begin{subfigure}[t]{0.32\textwidth}
        \centering
        \includegraphics[width=\linewidth]{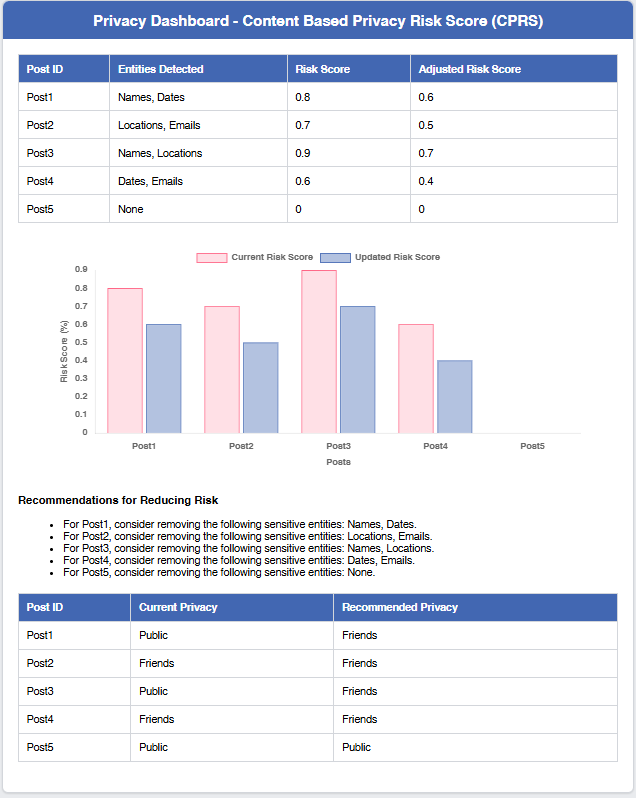}
        \caption{Content-Based Risk UI}
        \Description{Dashboard interface for highlighting sensitive content and content-based risk.}
        \label{fig:dashboard_content_based_privacy_risk_ui}
    \end{subfigure}
    \caption{Dashboard UI Screens: (a) Attribute-Based, (b) Social Graph-Based, and (c) Content-Based Privacy Risk Views.}
    \label{fig:dashboard_ui_all}
\end{figure*}

The user study addresses \textbf{RQ3 and RQ4}, evaluating CPRS with 100 participants of varying levels of privacy awareness. The cohort included undergraduate students, graduate students, and university staff (e.g., administrative, technical). Participants ranged in age from 18 to 50 years and represented a variety of academic departments, including Computer Science, Electrical and Computer Engineering (ECE), Business, and others. This demographic diversity enabled us to collect broad usability feedback across user roles, disciplines, and privacy mindsets. Each participant interacted with a simulated dashboard that visualized outputs from our unified CPRS model—including APRS, SGPRS, CBPRS, and the aggregated score—and received automated recommendations. Participants explored their risk profiles and rated clarity and usefulness on a 5-point Likert scale, also providing qualitative feedback. The Attribute-Based Privacy Risk page (Figure~\ref{fig:dashboard_attribute_risk_score_ui}) displays sensitivity and visibility risks for profile attributes like email or location and provides privacy suggestions. The Social Graph-Based Privacy Risk view (Figure~\ref{fig:dashboard_social_graph_based_privacy_risk_score_ui}) presents network-driven risk via structural similarity and importance, along with an interactive graph highlighting neighbor risks. The Content-Based Privacy Risk view (Figure~\ref{fig:dashboard_content_based_privacy_risk_ui}) identifies sensitive entities (e.g., names, dates, locations) in posts and comments, aggregates exposure, and offers content-level mitigation strategies.

\textbf{User Study Questions} Participants were asked to evaluate the dashboard through a set of seven Likert-scale questions (rated from 1 = not at all to 5 = extremely):
(1) How clear was the presentation of privacy risk scores for individual attributes in the dashboard?
(2) How useful were the recommendations for adjusting privacy settings based on the attribute risk scores?
(3) Did the dashboard make it easy to decide on changes to your privacy settings based on the adjusted privacy settings and scores for the attributes?
(4) How effectively did the dashboard highlight sensitive information from posts and comments?
(5) Were the recommendations for reducing content-based privacy risks actionable and easy to understand?
(6) How well did the dashboard help you understand a user's own risk score in the context of its social graph?
(7) Does seeing risk scores of neighbors of a user improve the understanding of its privacy exposure in the network?

Results show 85\% found risk presentation clear (50\% rated 5). Attribute recommendations were useful to 83\%; 82\% found the dashboard helpful for privacy decisions. Content explanations were effective for 78\%, with 74\% finding suggestions actionable. Social graph insights were helpful to 72\%, and 78\% said neighbor scores improved awareness. These results confirm CPRS’s effectiveness and usability.

\section{Ethical Consideration}
We follow strict ethical guidelines in using publicly available datasets such as SNAP and Koo, which were used solely for experimentation without redistribution or modification. User privacy is preserved through anonymization, with no personal identifiers stored—only aggregate results are reported. The user study followed an IRB-approved protocol, and all participants gave informed consent.

\section{Discussions and Limitations}
We addressed \textbf{RQ1} by demonstrating that privacy risks can be effectively quantified through the integration of attribute sensitivity and visibility (APRS), structural graph risk (SGPRS), and content-based exposure (CBPRS) within the CPRS framework (Section~\ref{methodology}). For \textbf{RQ2}, our results (Table~\ref{tab:privacy_risk_distribution}, Table~\ref{tab:cprs_results}) show that SGPRS has the highest average contribution (0.52), followed by CBPRS (0.48) and APRS (0.45), highlighting the dominant role of graph and content structures in overall risk. To answer \textbf{RQ3}, a user study (Section~\ref{user_study}) revealed that 83\%–85\% of participants found the CPRS-based risk insights and privacy recommendations actionable and useful for managing their settings. Finally, \textbf{RQ4} is supported by successful application of CPRS to real-world datasets (SNAP and Koo), along with high satisfaction scores in user evaluations (Section~\ref{user_study}), validating both the framework’s effectiveness and its practical utility.

Our CPRS framework integrates social graph and textual data for holistic privacy risk analysis but has scalability limitations for real-time use and depends on user-generated content quality. Platform-specific privacy settings require tuning, and uniform time-based sampling of 1M posts/comments may miss topic shifts or event-driven spikes. CPRS is both diagnostic and prescriptive—alerting users to risks and suggesting safer configurations. However, it currently lacks formal baseline comparisons. Future work will benchmark CPRS against simpler baselines like the visibility-only model of Liu and Terzi~\cite{LiuTerzi2009} to assess added value. Small-scale or controlled experiments may further clarify comparative performance. While CPRS scores are theoretically grounded, empirical validation is needed. Future strategies include testing CPRS against actual privacy breaches, adversarial inference, or controlled disclosures to assess correlation with real-world risk. Lastly, the user study provided useful feedback but had limited demographic diversity, suggesting the need for broader future evaluations.

\section{Conclusion}
In this paper we present a comprehensive framework (CPRS) for assessing privacy risks in social networks by unifying attribute sensitivity, social graph structure, and content-based exposure into a single scoring model. Validated on real-world datasets (SNAP and Koo), CPRS effectively quantifies user privacy risks and highlights the contribution of each dimension. Experiments demonstrate component-level impact on exposure, and a user study confirms the framework's usability and actionable insights. Future work includes incorporating multimodal content, temporal dynamics, and extending CPRS for real-time risk monitoring and adaptive interfaces to improve user control and transparency.


\begin{acks}
This work was supported in part by the U.S. National Science Foundation (Award No. 2451946) and the U.S. Nuclear Regulatory Commission (Award No. 31310025M0012). ChatGPT was utilized to assist with language editing and clarity improvements in this work. No content was generated related to technical results, data, code, or analysis.
\end{acks}

% \bibliographystyle{ACM-Reference-Format}
\bibliographystyle{ACM-Reference-Format}
\bibliography{references}

\appendix

\end{document}